\documentclass[11pt]{article}
\usepackage[sc]{mathpazo}
\usepackage{fullpage}
\usepackage{graphicx}
\usepackage{adjustbox}
\usepackage{multirow}
\usepackage{amsmath}
\usepackage[utf8]{inputenc}
\usepackage{lineno}
\usepackage{titlesec}
\usepackage{graphicx}
\usepackage{subcaption}
\usepackage{booktabs,multirow}
\usepackage{here}
\usepackage{url}
\usepackage{algorithmic}
\usepackage{algorithm}
\usepackage[numbers,sectionbib,sort]{natbib}  
\usepackage[T1]{fontenc}
\newcommand{\mat}[1]{\mbox{\boldmath{$#1$}}} 

\titleformat{\section}[block]{\Large\bfseries\filcenter}{\thesection}{1em}{}
\titleformat{\subsection}[block]{\Large\itshape\filcenter}{\thesubsection}{1em}{}
\titleformat{\subsubsection}[block]{\large\itshape}{\thesubsubsection}{1em}{}
\titleformat{\paragraph}[runin]{\itshape}{\theparagraph}{1em}{}[. ]

\title{Estimating the number of household TV profiles based in customer behaviour using Gaussian mixture model averaging}

\author{Gabriel R. Palma$^{1, 2,\ast}$ \and 
Sally McClean$^{3}$ \and
Brahim Allan$^{4}$ \and
Zeeshan Tariq$^{3}$ \and
Rafael A. Moral$^{1, 2}$}

\date{}

\begin{document}

\maketitle

\noindent{} 1. Hamilton Institute, Maynooth University, Maynooth, Ireland;

\noindent{} 2. Department of Mathematics and Statistics, Maynooth University, Maynooth, Ireland;

\noindent{} 3. Ulster University, Belfast, Northern Ireland;

\noindent{} 4. British Telecommunications, England

\noindent{} $\ast$ Corresponding author; e-mail: gabriel.palma.2022@mumail.ie

\bigskip

\bigskip

\textit{Keywords}: Bayesian modelling, Big data, Household TV profiles, Clustering, Dimensionality reduction.

\bigskip

\textit{Manuscript type}: Research paper.

\bigskip

\noindent{\footnotesize Prepared using the suggested \LaTeX{} template for \textit{Am.\ Nat.}}

\newpage{}

\section*{Abstract}
TV customers today face an abundance of choices from a growing number of live channels and on-demand services. Providing a personalised experience which saves customers time in discovering content is an essential requirement for TV providers, but a reliable understanding of their behaviour and preferences is key to achieving this. When trying to create personalised recommendations for TV, the biggest challenge is to understand viewing behaviour within households when multiple people are watching. The objective is to detect and combine individual profiles so that better personalised recommendations can be made for group viewing. Our challenge is that we have little explicit information about who is watching the devices at any time (either individuals or groups). Also, we do not have a way to combine more than one individual profile so that better recommendations can be made for group viewing. Here, we propose a novel framework using a Gaussian mixture model averaging to obtain point estimates for the number of household TV profiles, and a Bayesian random walk model to introduce uncertainty. We applied our approach using data from real customers whose TV-watching data totalled approximately half a million observations. We extracted several features, including channel transitions, program ratio, and session duration for customers over time. Moreover, we applied factor analysis to obtain latent variables in a lower dimension to observe the effect of this transformation on the estimation of the number of profiles. Our results indicate that combining our framework with the original features provides a means to estimate the number of household TV profiles and their characteristics, including shifts over time and a quantification of uncertainty. These insights can help media providers improve content personalisation and targeted advertising. Additionally, this study lays the foundation for integrating estimated profiles into recommendation engines, improving both individual and group viewing experiences.

\newpage{}

\section{Introduction}
User profiling is a well-reported problem in the literature that encompasses various applications, including the creation of personalised recommendation systems \cite{cufoglu2014, li2015, duric2023}. In telecommunications, multiple authors proposed methods to address this issue, including non-supervised learning-based methods, such as K-means, hierarchical clustering and others \cite{oh2012, zhang2015, li2017, moral2022, kumar2023}. Moreover, supervised learning-based methods, such as Support Vector Machines (SVM), Neural Networks (NNs), and Na\"ive Bayes algorithms, have been proposed for cases where the actual number of user profiles is known, i.e., the number of profiles in a household, whether representing a single individual or multiple users’ viewing behaviours \cite{zhu2023, kanwal2024}. However, modern-day customers exhibit diverse and dynamic behaviours, increasing the challenge and opportunities to develop new techniques to improve the state of the art of user profiling \cite{moral2022}. Also, the availability of a diverse data source that can describe the dynamic behaviour of modern customers introduces another layer of complexity to this problem. Some researchers have proposed the use of image-based systems for tailoring user profiling \cite{alam2020}; for example, the use of Convolutional Neural Networks (CNN) has been showing promising results as a technique to predict user recommendations \cite{dudekula2023}. Moreover, the use of set-top boxes (STB) data has been reported in the literature using Latent Dirichlet allocation (LDA), K-means, hierarchical clustering and other algorithms for user profiling \cite{chang2012, jiang2014, li2017}. 

Using STB with viewing data and no extra information about the user makes the user profiling problem even more challenging, bringing opportunities for implementing new techniques that assist the understanding of viewing data. \cite{kumar2023} highlights the importance of considering multiple algorithms when utilising cluster techniques, and these insights can be extended for user profiling based on STB data. Moreover, uncertainty estimation is still an open question for user profiling based on STB data. Therefore, the combination of methods considering the uncertainty of the number of profiles estimate brings a new perspective to the user profiling area in telecommunications. 

This paper presents a new approach for estimating the number of household TV profiles as well as uncertainty, based on TV watching behaviour. We used anonymised data collected from British Telecom (BT) customers between November 2020 and September 2021. The dataset contained information about the duration of programs watched and the type of channels viewed. Multiple profiles can be found in a given viewing behaviour dataset, represented by clusters of customers that have similar features based on their viewing behaviour. Beyond advancing user profiling techniques, this research has direct implications for businesses, particularly in the telecommunications and media sectors. Service providers such as BT can utilise the estimated household TV profiles to refine content recommendation strategies, enhance personalisation, and optimise targeted advertising. A deeper understanding of household viewing patterns enables businesses to improve customer engagement and retention by delivering more relevant and adaptive viewing experiences.

Furthermore, this study establishes a foundation for integrating such profiling techniques into real-world recommendation engines. By incorporating the estimated profiles into content delivery systems, service providers can enhance group-based recommendations, ensuring that suggested content aligns with the diverse preferences of multiple household members. Future research will focus on bridging this gap, exploring how these profiling insights can be operationalised within recommendation engines to deliver tangible benefits for both businesses and consumers. We aim to develop a new approach by combining Gaussian Mixture Models (GMM) to estimate household TV profiles with model averaging and a Bayesian random walk model to introduce uncertainty and temporal components to the modelling framework. The methods used in this paper contribute to user profiling by combining well-known frameworks to address the profiling issue for recommendation systems \cite{asabere2012}, serving as an initial study for further contributions to the area.

\section{Methods}
\subsection{Case study and feature engineering}
The YouView Data and Insights team initially provided an anonymised dataset containing customer TV-viewing data from BT YouView set-top boxes (STBs) to support our research. Informed consent was obtained for all participants. The data includes features from 228 customers recorded from November 2020 to September 2021, totalling approximately half a million observations. An observation includes information from a session, defined as the time a customer spends watching TV between turning on and off a STB. The information related to the channel's name being watched and the session duration was collected. In addition, the name of the program being watched was collected, including the ratio between the time where the customer watched the specific program in a session and the total session time. 

\begin{table}[ht]
\centering
\begin{tabular}{lrrrrr}
  \hline
  & \multicolumn{4}{c}{Samples} \\\cmidrule{3-5}
 Data source & Features & 1 & 2 & 3  \\
  \hline
  
  & Number & 10.00 & 9.00 & 4.00  \\ 
  Transitions & Channels & 3.00 & 3.00 & 3.00  \\ 
   & Absorbing states & 0.00 & 0.00 & 0.00  \\[0.2cm] 
  
  & Average & 0.32 & 0.58 & 0.54  \\ 
  & Median & 0.26 & 0.75 & 0.54  \\ 
  & Sd & 0.31 & 0.45 & 0.53  \\ 
  program & Skewness & 0.18 & -0.22 & 0.00  \\ 
  ratio & Kurtosis & 1.27 & 1.21 & 1.00  \\ 
  &  $2.5\%$ quantile & 0.00 & 0.01 & 0.08  \\ 
  & $97.5\%$ quantile & 0.71 & 1.00 & 1.00 \\[0.2cm] 
  
  & Average & 1861.10 & 7111.33 & 4201.00 \\ 
  & Median & 2321.00 & 7196.00 & 5499.00  \\ 
  & Sd & 1610.34 & 3753.26 & 2596.00  \\ 
  Session & Skewness & -0.22 & -0.90 & -1.15  \\ 
  duration & Kurtosis & 1.23 & 2.29 & 2.33  \\ 
  & $2.5\%$ quantile & 35.77 & 906.80 & 696.40  \\ 
  & $97.5\%$ quantile & 3465.00 & 10133.00 & 5499.00 \\
   \hline
\end{tabular}
\caption{Feature extracted from the watching behaviour of customers during a month.}
\label{ExtractedFeatures}
\end{table}

Based on these variables, we designed 17 new features obtained from session duration, program ratio (the ratio between the time that a customer watches a program and the total time of the program) and the transitions between channels watched by a customer in a STB. The obtained transition-based features were (i) the total number of channel transitions; (ii) the number of channels; and (iii) the number of channels that presented no transitions from them. We call (iii) absorbing states, given that once customers start watching this channel, they will not transition to another one for the entire session. Each feature summarises the customer activity over a month, and we used this information to estimate clusters. We presented a summary of these designed features in Table~\ref{ExtractedFeatures}.

To analyse the effect of dimensionality reduction on the household TV profile estimation using the proposed approach, we applied an Exploratory Factor Analysis (EFA) using the 17 designed features, reducing the dimensionality of the data to 4 latent variables based on eigenvalue analysis. Table~\ref{ExploratoryFactorAnalysis} present the latent variables produced by the EFA and the designed features. Also, we presented the EFA's loadings, showing the linear association among the latent variables and the 17 designed features. We carried out the EFA using the \texttt{lavaan} package \cite{lavaan}.

\subsection{Estimating the number of profiles}
We used a Gaussian Mixture Model (GMM) to estimate the number of household TV profiles per month. The probability density function for a GMM can be
written as:
$$f(x, \Psi) = \sum_{k=1}^{G} \pi_k \phi(x; \mat{\mu}_k, \mat{\Sigma}_k),$$
where $G$ is the number the number of mixture components, $\phi(\cdot)$ is the multivariate Gaussian probability density function, $\mat{\mu}_k$ is the vector of means and $\mat{\Sigma}_k$ is the variance-covariance matrix. 
When fitting the GMM, the optimal number of clusters can be chosen via information criteria. We used the Bayesian Information Criterion (BIC) to determine this. Since we may consider many different assumptions for the variance-covariance matrix $\mat{\Sigma}_k$, we used a model averaging approach to estimate the monthly number of TV profiles based on BIC. The estimate of the number of profiles is obtained by the BIC-weighted average of the GMM's number of mixture components. We used $14$ different variance-covariance structures as available in the \texttt{mclust} R package \cite{mclust}. For each variance-covariance structure, we calculated the BIC for the GMM with $1$ to $15$ mixture components, totalling $210$ BIC values, $\omega_{pg}$, which can be arranged in a matrix $\mat{\Omega}$, where $p \in \{1, 2, \ldots, 14\}$ and $g \in \{1, 2, \ldots, 15\}$. For all $\omega_{pg}$, we applied the function:
$$f(\omega_{pg}) = \frac{\exp \left( \frac{\omega_{pg} - \min(\mat{\Omega})}{2} \right)}{\sum_{p=1}^{14}\sum_{g=1}^{15} \exp \left( \frac{\omega_{pg} - \min(\mat{\Omega})}{2} \right)},$$
obtaining the matrix $\mat{\Omega}^{\prime}$ containing the weights $\omega^{\prime}_{pg}$. Finally, for a given vector of GMM mixture components $C = \{ 1, 2, \ldots, 15\}$, we estimated the number of household TV profiles per month by computing $\hat{G} = \sum_{p=1}^{14}\sum_{g=1}^{15} \omega^{\prime}_{pg} C_g$.

\subsection{Estimating uncertainty}
A Bayesian model was developed to estimate the uncertainty about the number of the household TV profiles estimates, $\hat{G}$, provided by the GMM averaging described in the previous section. To incorporate temporal dependence, we utilised a random walk formulation. We model the mean number of household TV profiles with a random intercept per household $i$. Since the estimated number of profiles for household $i$ at time $t$, $Y_{it}$, is always greater than or equal to 1, and is a continuous variable due to the model averaging approach, we propose the following hierarchical model:
\begin{align*}
Y_{it}|\beta_{i}, w_{it}, \xi_{it} &\sim \mbox{Truncated-Gamma}\left(\mu_{it}, \tau\right) \\
\log(\mu_{it}) &= \beta_{i} + w_{it} \\
w_{it} &= w_{i,t-1} + \xi_{it} \\
\beta_{i} &\sim \mathcal{N}(\beta_0, \sigma_{\beta}^2) \\
\xi_{it} &\sim \mathcal{N}(0, \sigma_{\xi}^2) \\[-0.5cm]
\end{align*}
where $\mu_{it}$ and $\tau$ are the mean and dispersion parameters of the gamma distribution (truncated from 1 to infinity), $\beta_i$ is a household-level random intercept, and $w_{it}$ is the random-walk process with Gaussian noise $\xi_{it}$.

We estimate the model using the Bayesian framework, with 10,000 MCMC iterations as burn-in, and 20,000 MCMC iterations with a thinning of 15, for each of four MCMC chains. We used vague priors for all unknown parameters, while respecting the parametric space \cite{gelman}. All prior distributions are specified below:
\begin{align*}
\beta_{0} &\sim \mathcal{N}(0, 100) \\
\sigma_{\beta}^2 &\sim \mathcal{U}(0, 100) \\
\sigma_{\xi}^2 &\sim \mathcal{U}(0, 100) \\
\tau &\sim \mathcal{U}(0, 50) \\
\end{align*}
Model implementation was carried out using JAGS \cite{plummer} within R \cite{rcore}. Although we do not provide the raw data to avoid intellectual property violations against BT or YouView, we provide all R code to implement the GMM averaging and Bayesian random ralk model at \url{https://github.com/GabrielRPalma/HouseholdTVProfileEstimation}. Finally, a summary of the proposed approach is described by the pseudo code~\ref{Algorithm}. 

\begin{algorithm}
\caption{TV Viewer Profiling and Uncertainty Estimation}
\begin{algorithmic}[1]
\STATE \textbf{Input:} Data with program ratio, channel transitions, session duration
\STATE \textbf{Output:} Estimated number of profiles, uncertainty measure

\STATE \textbf{Step 1: Feature Extraction}
\STATE Extract features based on:
\begin{itemize}
    \item program ratio
    \item Channel transitions
    \item Session duration
\end{itemize}

\STATE \textbf{Step 2: Apply Gaussian Mixture Model (GMM)}
\FOR{each number of clusters $k$ from 1 to 15}
    \STATE Fit GMM with $k$ clusters
    \STATE Compute BIC for the fitted model
\ENDFOR

\STATE \textbf{Step 3: Estimate Number of Profiles}
\STATE Use model averaging based on BIC values to estimate the number of profiles

\STATE \textbf{Step 4: Estimate Uncertainty}
\STATE Apply Bayesian Random Walk Model to estimate uncertainty

\STATE \textbf{Return:} Estimated number of profiles and uncertainty measure
\end{algorithmic}
\label{Algorithm}
\end{algorithm}

\begin{figure}
    \centering
    \includegraphics[width=0.8\linewidth]{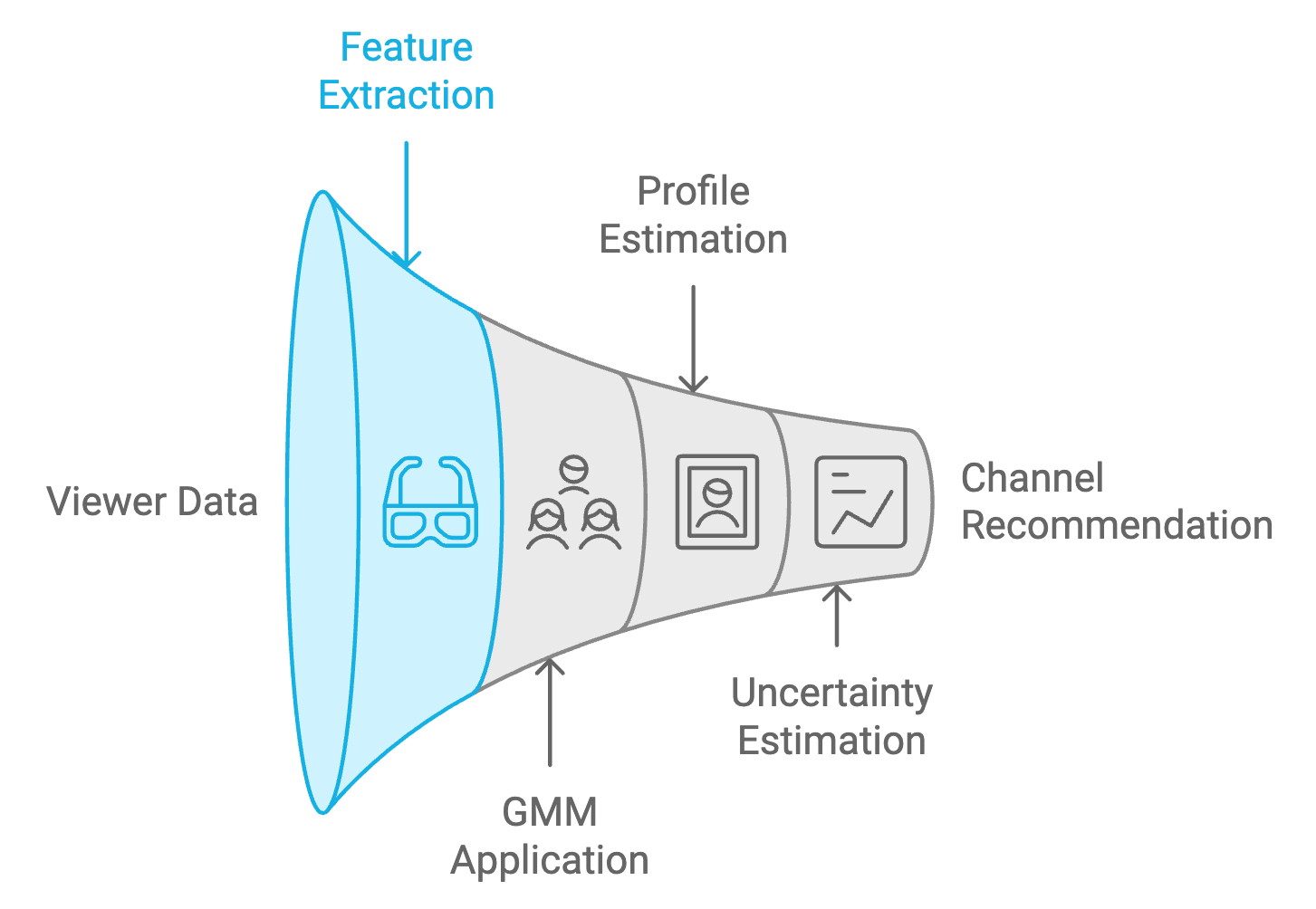}
    \caption{A diagram illustrating the complete proposed approach to estimate the number of household TV profiles and providing recommendations based on the following number.}
    \label{Diagram}
\end{figure}

\section{Results}
\subsection{Dimensionality reduction and cluster analysis}
After applying the exploratory factor analysis, the eigenvalue analysis indicated that four latent variables represent the overall features in a lower dimensional space containing $65.9\%$ of the data variability. The normalised loadings of these latent variables are presented in Table~\ref{ExploratoryFactorAnalysis}, indicating that the first latent variable emphasises features based on session duration, the second highlights position statistics of program ratio, and the third a combination of the three sources of features. Finally, the fourth latent variable highlights the dispersion features of the program ratio.

\begin{table}[ht]
\centering
\begin{tabular}{lrrrrr}
  \hline
  & \multicolumn{4}{c}{Latent Variables} \\\cmidrule{3-6}
 Data source & Features & 1 & 2 & 3 & 4  \\
  \hline
  
  & Number & - & - & $\mat{0.75}$ & $0.11$ \\  
  Transitions & Channels & - & $-0.31$ & $0.52$ & - \\ 
   & Absorbing states  & - & $-0.13$ & - & $-0.15$ \\[0.5cm]
  
  & Average  & $0.12$ & $\mat{0.64}$ & - & $0.48$ \\ 
  & Median  & $0.16$ & $\mat{0.78}$ & - & $0.22$ \\ 
  & Sd  & - & - & - & $\mat{0.94}$ \\ 
  program & Skewness & $-0.15$ & $-\mat{0.65}$ & $0.49$ &  -\\ 
  ratio & Kurtosis & - & - & $\mat{0.93}$ &  -\\ 
  &  $2.5\%$ quantile & - & $\mat{0.66}$ & $0.14$ & - \\ 
  & $97.5\%$ quantile & - & - & - & $\mat{0.94}$ \\[0.5cm] 
  
  & Average  & $\mat{0.84}$ & $0.28$ & - & - \\ 
  & Median  & $\mat{0.76}$ & $0.35$ & - & $-0.11$ \\ 
  & Sd  & $\mat{0.96}$ & $-0.24$ & - & $0.13$ \\ 
  Session & Skewness  & - & $-0.54$ & $0.44$  & $-0.20$ \\ 
  duration & Kurtosis  & - & $0.15$ & $\mat{0.78}$ & $-0.12$ \\ 
  & $2.5\%$ quantile & $0.20$ & $0.57$ & $0.15$ & - \\ 
  & $97.5\%$ quantile & $\mat{0.97}$ & - & - & - \\ 
   \hline
\end{tabular}
\caption{Normalised loading based on the explanatory factor analysis of the BT dataset. The bold values represent loadings greater than $0.6$.}
\label{ExploratoryFactorAnalysis}
\end{table}
We obtained an average of $9.16$ profiles, with a standard deviation ($sd$) of $3.93$ for the point estimates obtained by the GMM averaging using these latent variables considering all months and customers. On the other hand, by using all the features, we obtained an average of $4.87$ profiles ($sd = 3.96$). Also, the ratio of average distance within and between clusters obtained using the latent variables was $0.25$ ($sd = 0.16$), compared to $0.32$ ($sd = 0.21$) when using all features.

\begin{figure}
    \centering
    \includegraphics[width=0.8\linewidth]{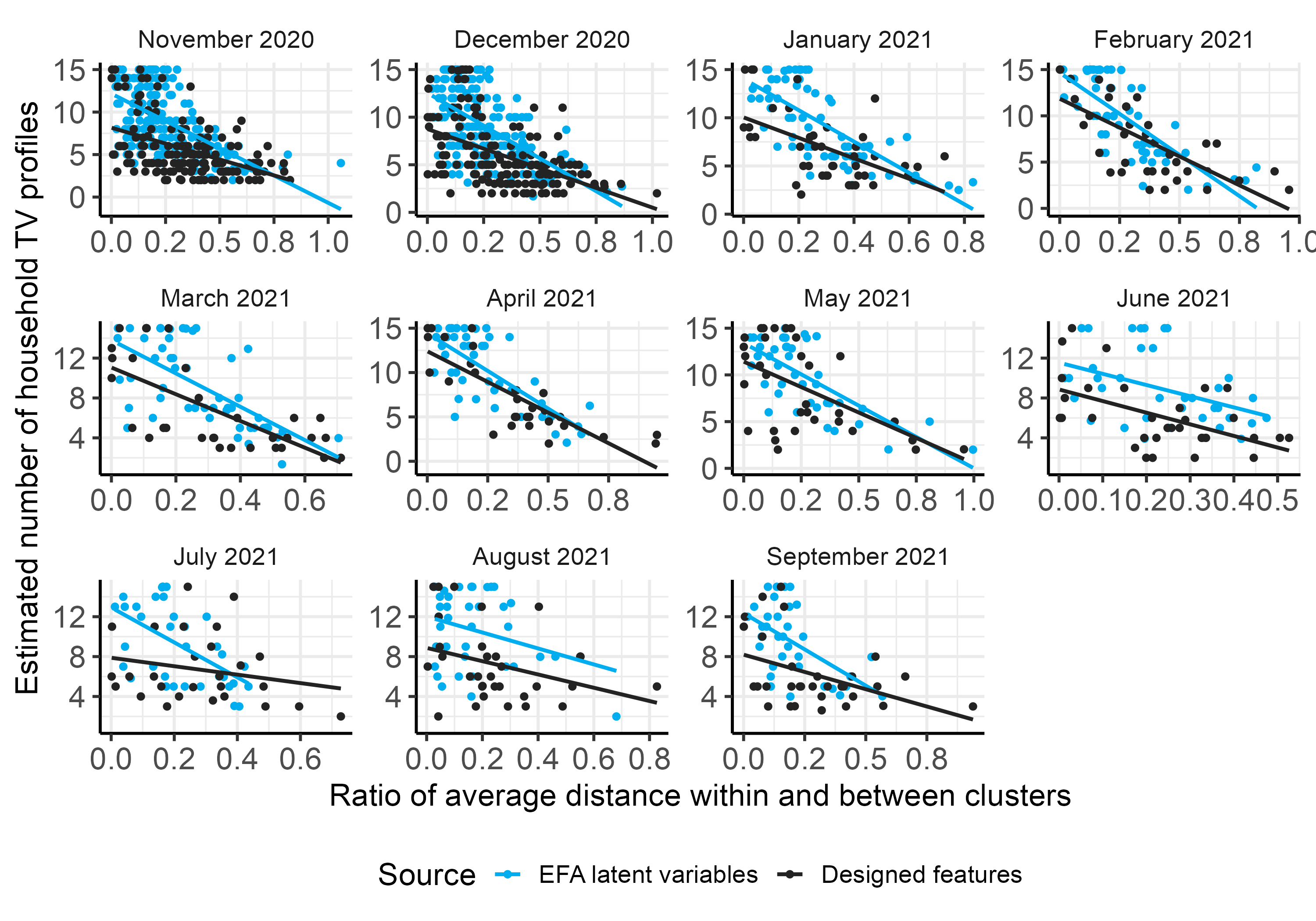}
    \caption{Scatter plot of point estimates of household TV profiles using Gaussian Mixture Models averaging and the ratio of average distance within and between clusters obtained from the Exploratory Factor Analysis latent variables and all features of watching behaviour.}
    \label{PointEstimateVSRatioDistance}
\end{figure}

Figure~\ref{PointEstimateVSRatioDistance} summarises the relationship between the point estimates of household TV profiles and the ratio of average distance within and between clusters for the months evaluated in this study. Fewer household TV profiles tend to have higher values of the ratio of average distance within and between clusters for all months and data sources used. This suggests that the clusters are less compact and not well separated when the number of household TV profiles is lower.

\begin{figure}
    \centering
    \includegraphics[width=0.8\linewidth]{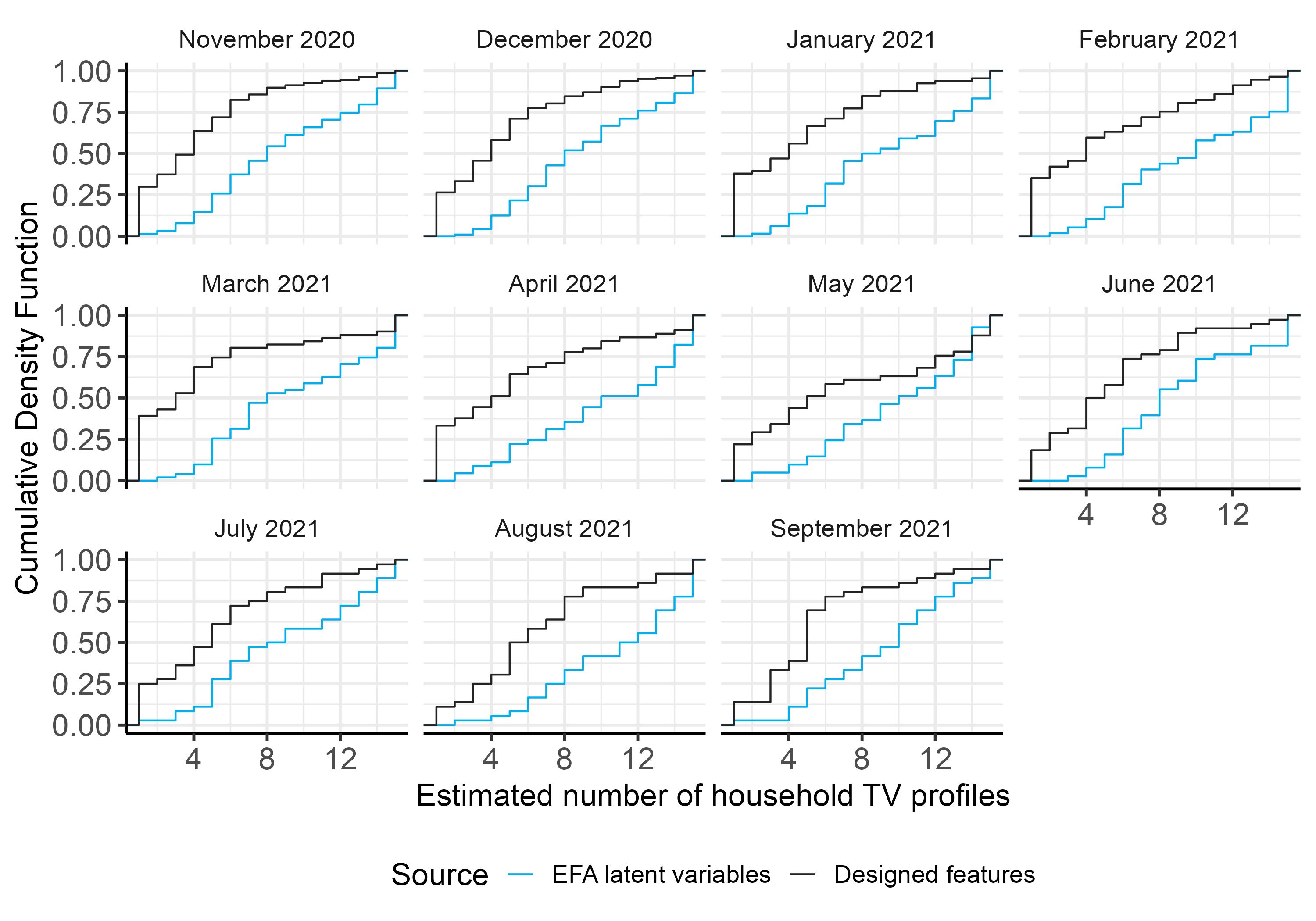}
    \caption{Cumulative density function of the estimated household TV profiles over ten months (December 2020 to September 2021) for all customers included in our dataset.}
    \label{CummulativeDistribution}
\end{figure}

Finally, Figure~\ref{CummulativeDistribution} highlights the obtained profiles based on the GMM and shows that the dimension reduction increased the number of estimated household TV profiles in all months analysed in our study. 

\subsection{Introducing uncertainty via the Bayesian hierarchical model}
\begin{figure}
    \centering
    \includegraphics[width=0.8\linewidth]{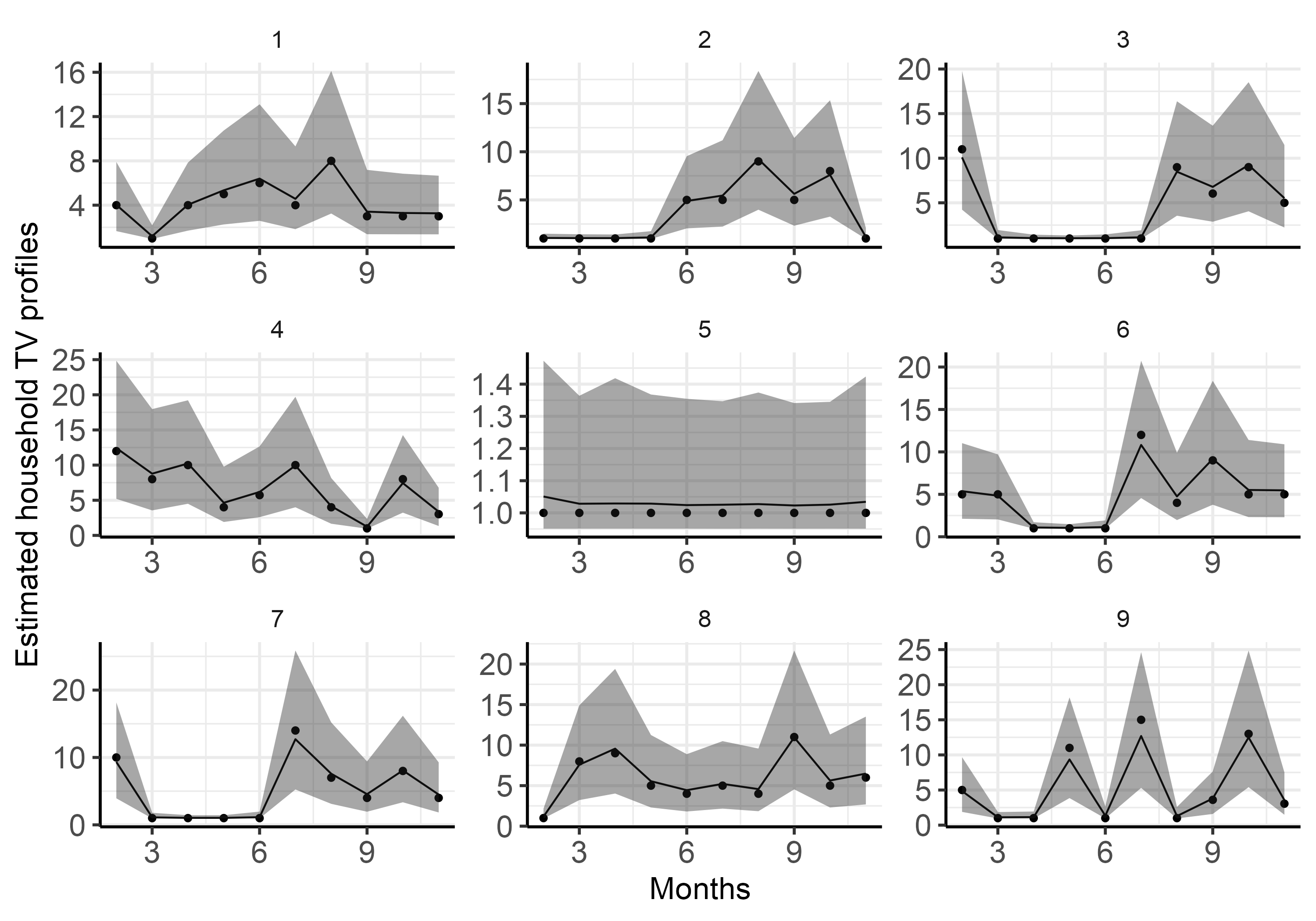}
    \caption{Time series of household TV profiles over ten months (December 2020 to September 2021) for nine customers, where lines represent the posterior means alongside the 95\% credible intervals as the gray ribbons. The points represent the point estimates provided by the Gaussian mixture model averaging.}
    \label{PredictionsBayesianModel}
\end{figure}

We fitted the Bayesian random walk model to the STB dataset for customers, where their viewing behaviour was recorded during all months. Figure~\ref{PredictionsBayesianModel} presents nine customers' estimates of the number of household TV profiles over time with their 95\% credible intervals for the proposed approach using all 17 features extracted from the original dataset that are presented in Table~\ref{ExtractedFeatures}.

The model performs well when there are few household TV profiles by incorporating uncertainty in scenarios where the variability of household TV profiles is smaller over the months, such as customer $5$ presented in Figure~\ref{PredictionsBayesianModel}. Also, the uncertainty increases in scenarios where the variability of the number of household TV profiles increases over time. Given that the ground truth for the number of household TV profiles is not available for this problem, the incorporated uncertainty is a better alternative than solely relying on the point estimates from the GMM averaging.

\section{Discussion}
We proposed a novel approach to estimate household TV profiles using Gaussian Mixture Model (GMM) averaging and a Bayesian random walk model. The approach can identify temporal patterns of household TV profiles, and GMM allows us to obtain metrics of goodness-of-fit and quality of specific clusters. Also, the proposed Bayesian model can overcome missing data by estimating the overall trend based on multiple customers. Thus, profiles can be estimated based on global watching behaviour patterns. We analysed the possibility of applying factor analysis to reduce the dimension of the dataset. Our study showed that by using this technique, the estimates of the number of profiles increased. However, cluster performance is slightly reduced, as indicated by the ratio of the mean distance within and between clusters.

Our results showed promising results for estimating the number of household TV profiles given the performance metrics of the GMM clusters. The proposed approach introduces a novel combination of well-established methods to estimate the number of household TV profiles, and it serves as an initial study towards the creation of recommendation systems that utilise the information of the estimated clusters' characteristics for a better customer TV experience. The proposed framework has significant implications for businesses, particularly for telecommunications and media providers such as BT. By accurately estimating the number of household TV profiles, service providers can refine content personalisation, leading to improved user experience and customer satisfaction. This approach can be integrated into existing recommendation systems to tailor content suggestions not only for individual viewers but also for shared household viewing, addressing a key challenge in TV analytics. Furthermore, by identifying shifts in viewing patterns over time, businesses can refine their marketing strategies, optimise content licensing decisions, and improve the targeting of advertisements. Given BT’s investment in innovation through initiatives such as BTIIC, incorporating such data-driven methods into content delivery platforms can offer a competitive advantage in an increasingly fragmented media landscape.

In future work, we will explore the characterisation of the clusters produced by the GMM averaging further by linking the extracted features to each cluster, such as the average program-watching profile for different clusters. Also, the assumption that the number of household TV profiles, $Y_{it}$, per customer $i$, varies every month $t$ follows a random walk can be relaxed in future research by incorporating, for example, weekend effects and seasonal behaviour. 

We identify two main limitations to this study. First, there was no ground truth in the available dataset to evaluate the accuracy of the estimated profiles. However, our findings serve as proof of concept by presenting the cluster performances and evaluating the effect of the latent variables in the number of estimated household TV profiles based on the proposed approach. Finally, our analysis was limited to monthly estimations of household TV profiles, considering the limited amount of data used in this initial study. However, the approach presented is a proof of concept and can be applied for daily estimates if there is more data availability. We emphasise the heterogeneity of customers' watching behaviour, which can impact the estimation of the number of profiles. Therefore, additional steps are necessary to enhance the practical benefits of the estimated profiles towards recommendation systems. Future work may be carried out to address these limitations by (1) obtaining a new dataset with the actual number of household TV profiles so that the performance of the proposed approach can be tested, (2) obtaining more data to achieve daily estimates of household TV profiles, (3) create an additional step in the proposed framework to deal with missing data in cases that customers were not monitored in all periods evaluated in the study for the estimation of the number of household TV profiles, and (4) integrating the estimated household TV profiles into a recommendation engine would enable service providers to deliver tailored content suggestions at both individual and household levels. This would allow households to benefit from more relevant and engaging TV experiences, leveraging the advantages of this profiling approach.

\section{Acknowledgments}

This publication has resulted from research conducted with the support of BTIIC (the BT Ireland Innovation Centre), funded by BT, Invest Northern Ireland and Taighde Éireann – Research Ireland under Grant 18/CRT/6049. We are also thankful to the YouView team for providing the sample data and domain knowledge. The opinions, findings and conclusions or recommendations expressed in this material are those of the authors and do not necessarily reflect the views of the funding agencies. 

\section{Declarations}
~~~~
\textbf{Ethical Approval} Not applicable.

\textbf{Competing interests} Not applicable.


\begin{thebibliography}{00}

\bibitem{vanderaalst} W. M. Van der Aalst, ``Process modeling and analysis,'' in \emph{Process mining: Data science in action,} 2\textsuperscript{nd} ed. Berlin, Germany: Springer, 2016, ch. 3, pp. 55--88.

\bibitem{amato} F. Amato, A. Castiglione, A. De Santo, V. Moscato, A. Picariello, F. Persia, and G. Sperl\'i ``Recognizing human behaviours in online social networks,'' \emph{Computers \& Security,} vol. 74, pp. 355--370, 2018

\bibitem{plummer} M. Plummer ``JAGS: A program for analysis of Bayesian graphical models using Gibbs sampling,'' \emph{Proceedings of the 3rd international workshop on distributed statistical computing,} vol. 124. no. 125.10, pp. 1--10, 2003.

\bibitem{rcore} R Core Team, ``R: A language and environment for statistical computing," The R Foundation for Statistical Computing, Vienna, Austria.

\bibitem{nelder} P. McCullagh, J.A. Nelder, ``Generalized Linear Models," in \emph{Imprint Routledge,} 2 edition, 1989

\bibitem{chen} Z. Chen, S. Zhang, S. McClean, B. Allan, B., and I. Kegel. ``Sequence Mining TV Viewing Data Using Embedded Markov Modelling.'' \emph{2021 IEEE SmartWorld, Ubiquitous Intelligence \& Computing, Advanced \& Trusted Computing, Scalable Computing \& Communications, Internet of People and Smart City Innovation (SmartWorld/SCALCOM/UIC/ATC/IOP/SCI)}, pp. 665--670, 2021.

\bibitem{lee2023} H. Lee, ``Uncovering and understanding smart TV users’ picture quality preferences via big data analytics,'' in \emph{2023 IEEE International Conference on Big Data (BigData)}, IEEE, 2023, pp. 1764--1773.

\bibitem{jiang2014} X. Jiang, Y. Wang, and J. Chai, ``Research of users' viewing habits based on clustering method,'' in \emph{2014 IEEE 3rd International Conference on Cloud Computing and Intelligence Systems}, IEEE, 2014, pp. 331--335.

\bibitem{chang2012} R. M. Chang, R. J. Kauffman, and I. Son, ``Consumer micro-behavior and TV viewership patterns: data analytics for the two-way set-top box,'' in \emph{Proceedings of the 14th Annual International Conference on Electronic Commerce}, 2012, pp. 272--273. 

\bibitem{li2015} H. Li et al., ``Personalized TV recommendation with mixture probabilistic matrix factorization,'' in \emph{Proceedings of the 2015 SIAM International Conference on Data Mining}, Society for Industrial and Applied Mathematics, 2015, pp. 352--360.

\bibitem{duric2023} I. Đuric et al., ``Model of an intelligent smart home system based on ambient intelligence and user profiling,'' \emph{Journal of Ambient Intelligence and Humanized Computing}, vol. 14, no. 5, pp. 5137--5149, 2023.

\bibitem{kumar2023} R. Kumar et al., ``Clustering the various categorical data: An exploration of algorithms and performance analysis,'' in \emph{2023 4th International Conference for Emerging Technology (INCET)}, IEEE, 2023, pp. 1--6.

\bibitem{li2017} Z. Li, R. Kauffman, and B. Dai, ``Can I see beyond what you see? Blending machine learning and econometrics to discover household TV viewing preferences,'' 2017.

\bibitem{zhang2015} H. Zhang et al., ``Application of clustering algorithm on TV programs preference grouping of subscribers,'' in \emph{2015 IEEE International Conference on Computer and Communications (ICCC)}, IEEE, 2015, pp. 40--44.

\bibitem{oh2012} J. Oh et al., ``Time-dependent user profiling for TV recommendation,'' in \emph{2012 Second International Conference on Cloud and Green Computing}, IEEE, 2012, pp. 783--787.

\bibitem{zhu2023} Y. Zhu, Z. Li, L. Sun, and L. Gao, ``Lightweight multi-role recommendation system in TV live-streaming,'' in \emph{2023 IEEE International Symposium on Broadband Multimedia Systems and Broadcasting (BMSB)}, Beijing, China, 2023, pp. 1--6. DOI: 10.1109/BMSB58369.2023.10211234.

\bibitem{dudekula2023} K. V. Dudekula et al., ``Convolutional neural network-based personalized program recommendation system for smart television users,'' \emph{Sustainability}, vol. 15, no. 3, p. 2206, 2023.

\bibitem{asabere2012} N. Y. Asabere, ``A survey of personalized television and video recommender systems and techniques,'' \emph{Inf. Commun. Technol. Res.}, vol. 2, pp. 602--608, 2012.

\bibitem{kanwal2024} M. Kanwal, N. A. Khan, and A. A. Khan, ``A machine learning approach to user profiling for data annotation of online behavior,'' \emph{Computers, Materials \& Continua}, vol. 78, no. 2, pp. 2721--2739, 2024.

\bibitem{alam2020} I. Alam and S. Khusro, ``Tailoring recommendations to groups of viewers on smart TV: a real-time profile generation approach,'' \emph{IEEE Access}, vol. 8, pp. 50814--50827, 2020.

\bibitem{cufoglu2014} A. Cufoglu, ``User profiling-a short review,'' \emph{International Journal of Computer Applications}, vol. 108, no. 3, pp. 1--9, 2014.

\bibitem{moral2022} R. A. Moral et al., ``Profiling television watching behavior using Bayesian hierarchical joint models for time-to-event and count data,'' \emph{IEEE Access}, vol. 10, pp. 113018--113027, 2022.

\bibitem{mclust} L. Scrucca, C. Fraley, TB. Murphy, AE. Raftery (2023). Model-Based Clustering, Classification, and Density Estimation Using mclust in R. Chapman and Hall/CRC. ISBN 978-1032234953, doi:10.1201/9781003277965, https://mclust-org.github.io/book/.

\bibitem{lavaan} Rosseel Y (2012). “lavaan: An R Package for Structural Equation Modeling.” Journal of Statistical Software, 48(2), 1–36. doi:10.18637/jss.v048.i02.

\bibitem{gelman} A. Gelman, J.B. Carlin, H.S. Stern, D.B. Dunson, A. Vehtari, D.B. Rubin (2013). Bayesian Data Analysis. Chapman and Hall/CRC, 3rd edition. ISBN 978-1439840955


\end{thebibliography}
\end{document}